\documentstyle[11pt]{article}
\normalbaselines
\begin{document}

\begin{center}

{\LARGE\bf Quasi-exactly solvable problems\\[0.5cm]
and random matrix theory}

\vskip 1cm

{\large {\bf G.M. Cicuta} }

\vskip 0.1 cm

Dipartimento di Fisica, Universita di Parma,\\
Viale delle Scienze, I - 43100 Parma, Italy \footnote{E-mail
address: cicuta@parma.infn.it} \\[0.5cm]

and\\[0.5cm]

{\large {\bf A.G. Ushveridze} }\footnote{This work was
partially supported by the Deutsche Forschungs Gemainschaft
(DFG) and the Lodz University grant no. 457}

\vskip 0.1 cm

Department of Theoretical Physics, University of Lodz,\\
Pomorska 149/153, 90-236 Lodz, Poland\footnote{E-mail
address: alexush@mvii.uni.lodz.pl and alexush@krysia.uni.lodz.pl} \\

\end{center}
\vspace{1 cm}
\begin{abstract}

There exists an exact relationship
between the quasi-exactly solvable problems of quantum
mechanics and models of square and rectangular random
complex matrices. This relationship enables one to reduce the
problem of constructing topological ($1/N$) expansions
in random matrix models to the problem of constructing semiclassical expansions
for observables in quasi-exactly solvable problems.  Lie algebraic
aspects of this relationship are also discussed.

\end{abstract}

\newpage

\section{Introduction. The idea}

The method of orthogonal polynomials \cite{bessis} is one of the basic
tools for doing calculations in the theory of large random
matrices (see e.g. recent review papers \cite{alvarez,jerus,lozano}).
In order to remind the reader the idea of this method, let
us consider for definiteness a model of random hermitean
matrices $\Phi_N$ of size $(N+1)\times (N+1)$ described by the
potential $V(\Phi_N)$. The `observables' in this model are
initially defined as fractions of multiple integrals
\begin{eqnarray}
\langle f \rangle=\frac{\int D\Phi_N \mbox{Tr}\
f(\Phi_N)\exp\{-\mbox{Tr}\ V(\Phi_N)\}}
{\int D\Phi_N \exp\{-\mbox{Tr}\ V(\Phi_N)\}}
\label{1}
\end{eqnarray}
over components of matrix $\Phi_N$. The measure of the integration,
$D\Phi_N$, as well as subintegral expressions are invariant
under unitary transformations, $\Phi_N\rightarrow U_N\Phi_N
U_N^{-1}$, and this fact enables one to extract from both
integrals (and then cancel) the volume of the unitary group
by means of the well known Faddeev -- Popov method.
After this, formula (\ref{1}) takes simpler form
\begin{eqnarray}
\langle f \rangle=\frac{\int d\lambda_0\ldots d\lambda_N \Delta^2(\lambda)
\sum_{i=0}^N f(\lambda_i)\exp\{-\sum_{i=0}^N V(\lambda_i)\}}
{\int d\lambda_0\ldots d\lambda_N \Delta^2(\lambda)
\exp\{-\sum_{i=0}^N V(\lambda_i)\}}
\label{2}
\end{eqnarray}
in which the variables $\lambda_0,\ldots,\lambda_N$ have the meaning of
eigenvalues of matrix $\Phi_N$, and $\Delta_N(\lambda)=\det
||\lambda_i^n||$ is the Vandermonde determinant of these eigenvalues.

A brilliant idea by Bessis \cite{bessis} was
to represent this determinant in the form $\Delta_N(\lambda)=\det
||P_n(\lambda_i)||$, where $P_n(\lambda)$ are polynomials
normalized as $P_n(\lambda)=\lambda^n+\ldots$ and
orthogonal with the weight $e^{-V(\lambda)}$. This
replacement enables one to reduce formula (\ref{2}) to the
form
\begin{eqnarray}
\langle f \rangle=\sum_{n=0}^N f_{nn}
\label{3}
\end{eqnarray}
where
\begin{eqnarray}
f_{nn}=\frac{\int dx\ f(x) \Psi_n^2(x)}{\int dx\ \Psi_n^2(x)}
\label{4}
\end{eqnarray}
and
\begin{eqnarray}
\Psi_n(x)=P_n(x)e^{-V(x)/2}.
\label{5}
\end{eqnarray}
{}From the standard theory of orthogonal polynomials (see,
e.g. \cite{szego}) it
follows that, if function $f(x)$ is a polynomial of degree $k$, then
the quantities $f_{nn}$ can be found from simple
recurrence relations of the depth $k+2$. In large $N$ limit
these recurrence relations can be reduced to a differential equation
whose solution determines the leading term of the topological ($1/N$)
expansion for (\ref{1}). The construction of the next terms
of this topological expansion is generally rather complicated
matter, because the $n$-dependence of numbers $f_{nn}$ is not
always smooth \cite{molinari}.
The idea of the approach which we intend to present in this
paper is to find such random matrix models for which
the functions (\ref{5}) could be identified with
orthogonal wavefunctions for some quantum mechanical model.
In this case the quantity $\langle f \rangle$ would have
the meaning of the trace of function $f(x)$ over the first $N$ excitations
in this model. In large $N$ limit, this trace could easily be computed (with
an arbitrary accuracy) in the framework of WKB
approximation. In this case the computation of several
first terms of the topological expansion for (\ref{1})
should not encounter serious difficulties.

Of course, such an identification is not always possible,
because the wavefunctions of quantum mechanical models are
not obliged to have the form (\ref{5}). There are however,
two classes of models for which the form of wavefunctions is
given by formula (\ref{5}). These are {\it exactly solvable
models} associated with classical orthogonal polynomials
(see e.g. \cite{flugge})
and the so-called {\it quasi-exactly solvable models}
associated with non-standard orthogonal polynomials \cite{ushbook}.
The corresponding random matrix models we respectively
shall call {\it exactly}- and {\it quasi-exactly solvable random
matrix models}.

It is not difficult to see that the exactly solvable random
matrix models are really exactly solvable in the standard
sense of this word. Consider, for example, the simplest case
of the harmonic oscillator with the potential $W(x)=x^2$. The
wavefunctions in this model have the form $\Psi_n(x)=H_n(x)e^{-x^2/2}$,
where $H_n(x)$ are Hermite polynomials. Comparing this form
with (\ref{5}) we find the form of the random matrix potential:
$V(x)=x^2$. But this potential describes the Gauss ensemble of
random matrices and is exactly solvable \cite{metha}.
It seems possible that other
exactly solvable quantum models like Morse potential, P\"oschel--Teller
potential well, etc. \cite{flugge}, associated with other classical
orthogonal polynomials, will be related to
random matrix models exactly solvable for any N.

Much more non-trivial and interesting classes of random
matrix models appear if one starts with the quasi-exactly
solvable models in quantum mechanics \cite{ushbook}. Remember that these
models, which have been discovered several years ago
\cite{zasl,bagr,turush,ush,tur}, are
distinguished by the fact that they can be solved exactly (analytically)
only for a finite number of
eigenvalues and corresponding eigenfunctions. It is however
remarkable that the number $N$ of exactly
calculable energy levels is a free parameter
of the hamiltonian and can be chosen arbitrarily.
The aim of the present paper is to describe the
quasi-exactly solvable random matrix models associated with
some simplest quasi-exactly solvable quantum mechanical problems.

\section{Quasi-exactly solvable sextic oscillator model}

Consider a one-dimensional sextic anharmonic
oscillator with hamiltonian
\begin{eqnarray}
H_N=-\hbar_N^2\frac{\partial^2}{\partial x^2}+
\{b^2-[8-2\hbar_N]a\}x^2+4abx^4+4a^2x^6,
\label{a.1}
\end{eqnarray}
in which $a>0$ and $b$ are real parameters and $N$ is an
arbitrarily chosen non-negative integer. The parameter
$\hbar_N$ is defined as $\hbar_N\equiv (N+1)^{-1}$ and has
the meaning of the ``quantized'' Planck constant.
It is not difficult to show that the model (\ref{a.1}) is
{\it quasi-exactly solvable} and has the {\it order} $N+1$ \,\cite{ushbook}.
This means that  Schr\"odinger equation for
(\ref{a.1}) for any given $N$ admits exact (algebraic) solutions only
for $N+1$ energy
levels and corresponding wavefunctions. An explicit form
of these solutions is given by the formulas
\begin{eqnarray}
E_n=(4-3\hbar_N)b+8a\hbar_N\sum_{i=1}^N \xi_{n,i},
\qquad n=0,\ldots, N
\label{a.2}
\end{eqnarray}
and
\begin{eqnarray}
\Psi_n(x)=c_n\prod_{i=1}^N \left(x^2-\xi_{n,i}\right)\exp
\left[-\frac{ax^4+bx^2}{2\hbar_N}\right],
\qquad n=0,\ldots, N,
\label{a.3}
\end{eqnarray}
in which $\{\xi_{n,1},\dots, \xi_{n,N} \},\ n=0,\ldots, N$ are
$N+1$ different sets of real numbers satisfying
the system of $N$ algebraic equations
\begin{eqnarray}
\sum_{k=1,k\neq i}^N\frac{\hbar_N}{\xi_{n,i}-\xi_{n,k}}+
\frac{\hbar_N}{4\xi_{n,i}}-
\frac{b}{2}-a\xi_{n,i}=0, \qquad i=1,\ldots N.
\label{a.4}
\end{eqnarray}
It can be shown that $N+1$ solutions
of system (\ref{a.4}) describe the first $N+1$ even energy
levels in model (\ref{a.1}), i.e. levels with numbers $2n, \ \ n=0,\ldots,
2N$. The linear span of corresponding
wavefunctions forms a $N+1$-dimensional subspace of
Hilbert space which we denote by ${\cal H}_N$.

\section{Non-standard orthogonal polynomials}

We choose the normalization constants $c_n$ in formula (\ref{a.3})
in such a way as to guarantee the
orthonormalizability of wave functions $\Psi_n(x)$:
\begin{eqnarray}
\int_{-\infty}^{+\infty}\Psi_n(x)\Psi_m(x)dx=\delta_{nm}.
\label{a.5}
\end{eqnarray}
Representing wavefunctions in the form
\begin{eqnarray}
\Psi_n(x)={\cal P}_n(x^2)\exp\left\{-\frac{ax^4+bx^2}{2\hbar_N}\right\}
\label{a.6}
\end{eqnarray}
where
\begin{eqnarray}
{\cal P}_n(x^2)=c_n\prod_{i=1}^N \left(x^2-\xi_{n,i}\right),
\label{a.8}
\end{eqnarray}
we can rewrite (\ref{a.5}) as
\begin{eqnarray}
\int_{-\infty}^{+\infty} dx {\cal P}_n(x^2){\cal P}_m(x^2)
\exp\left\{-\frac{ax^4+bx^2}
{\hbar_N}\right\}=\delta_{nm}.
\label{a.9}
\end{eqnarray}
{}From formula (\ref{a.9}) it follows
that ${\cal P}_n(x^2)$ can be considered as polynomials
orthogonal with the weight $\exp[-(ax^4+bx^2)/\hbar_N]$.
Of course, ${\cal P}_n(x^2)$ are not the classical orthogonal
polynomials because they are of the same degree $N$.
The general form of these polynomials is
\begin{eqnarray}
{\cal P}_n(x^2)=\sum_{m=0}^N P_{nm}x^{2m}, \qquad
n=0,\ldots, N,
\label{a.10}
\end{eqnarray}
where $P_{nm}$ is a certain non-degenerate (and non-triangular!)
$(N+1)\times (N+1)$ matrix.

It is worth stressing that in
the theory of random matrices a basic tool is provided by the set
of monic polynomials ($ P_n= x^n $ + lower degree monomials) which are
orthogonal
\begin{eqnarray}
\int_{-\infty}^{+\infty} dx P_n(x^2) P_m(x^2)
\exp\left\{-\frac{ax^4+bx^2}
{\hbar_N}\right\}= h_n \, \delta_{nm}.
\label{b.1}
\end{eqnarray}
They obey the recursion relation
\begin{eqnarray}
x P_n(x)= P_{n+1} + R_n P_{n-1}
\label{b.2}
\end{eqnarray}
and may be considered completely known, since Bessis \cite{bessis}
evaluated the generating function of the momenta $\mu_k$
\begin{eqnarray}
\mu_k =\int_{-\infty}^{+\infty} dx \, x^k
\exp\left\{-\frac{ax^4+bx^2}{\hbar_N}\right\}
\label{b.3}
\end{eqnarray}
and related the coefficients $R_n$ to the momenta.

\section{From quantum mechanics to random matrix theory}

Let us now take an arbitrary even function $f(x^2)$ and consider
its trace $\mbox{Tr}_N f(x^2)$ in the space ${\cal H}_N$, i.e. in the space of
all exactly calculable wavefunctions:
\begin{eqnarray}
\mbox{Tr}_N f(x^2)=\sum_{n=0}^N \int_{-\infty}^{+\infty}
f(x^2)\Psi_n^2(x)dx.
\label{a.16}
\end{eqnarray}
Using (\ref{a.6}) and (\ref{a.9}), we can rewrite
(\ref{a.16}) as
\begin{eqnarray}
\mbox{Tr}_N f(x^2)=\sum_{n=0}^N \frac
{\int_{-\infty}^{+\infty} dx f(x^2) {\cal P}_n^2(x^2)\exp[-\hbar_N^{-1}
(ax^4+bx^2)]}{\int_{-\infty}^{+\infty} dx {\cal P}_n^2(x^2)\exp[-\hbar_N^{-1}
(ax^4+bx^2)]}.
\label{a.11}
\end{eqnarray}
It is not difficult to see that this expression can also be
represented in the form of a fraction of multiple integrals
\begin{eqnarray}
\mbox{Tr}_N f(x^2)=\frac
{\int_{-\infty}^{+\infty} \prod_{i=1}^{N+1}dx_i \left\{\sum_{i=1}^{N+1}
f(x_i^2)\right\} \det
||{\cal P}_n(x^2_i)||^2
\exp\left\{-\hbar_N^{-1}\sum_{i=1}^{N+1} (ax^4_i+bx^2_i)\right\}}
{\int_{-\infty}^{+\infty} \prod_{i=1}^{N+1}dx_i \det
|| {\cal P}_n(x^2_i)||^2
\exp\left\{-\hbar_N^{-1}\sum_{i=1}^{N+1} (ax^4_i+bx^2_i)\right\}}.
\label{a.12}
\end{eqnarray}
Using formula (\ref{a.10}), we can write
\begin{eqnarray}
\det ||{\cal P}_n(x^2_i)||=\det ||P_{nm}x_i^{2m}||=\det
||P_{nm}||\cdot \det ||x_i^{2m}||=
\det||P_{nm}||\cdot
\prod_{i<k}^{N+1}(x_i^2-x_k^2),
\label{a.13}
\end{eqnarray}
after which formula (\ref{a.12}) takes the form
\begin{eqnarray}
\mbox{Tr}_N f(x^2)=
\frac{\int_{-\infty}^{+\infty} \prod_{i=1}^{N+1}dx_i \left\{\sum_{i=1}^{N+1}
f(x_i^2)\right\} \prod_{i\neq k}^{N+1}(x_i^2-x_k^2)
\exp\left\{-\hbar_N^{-1}\sum_{i=1}^{N+1} (ax^4_i+bx^2_i)\right\}}
{\int_{-\infty}^{+\infty} \prod_{i=1}^{N+1}dx_i
\prod_{i\neq k}^{N+1}(x_i^2-x_k^2)
\exp\left\{-\hbar_N^{-1}\sum_{i=1}^{N+1} (ax^4_i+bx^2_i)\right\}}.
\label{a.14}
\end{eqnarray}
But from the random matrix theory we know that
this expression can be rewritten as
\begin{eqnarray}
\mbox{Tr}_N f(x^2)=
\frac{\int D[\Phi_N ^{\dag},\Phi_N] \left\{\mbox{Tr}
f(\Phi_N ^{\dag} \Phi_N)\right\}
\exp\left\{-\frac{1}{\hbar_N}\mbox{Tr}
\left[a(\Phi_N ^{\dag}    \Phi_N)^2+b \Phi_N ^{\dag} \Phi_N\right]\right\}}
{\int D[\Phi_N ^{\dag} ,\Phi_N]
\exp\left\{-\frac{1}{\hbar_N}\mbox{Tr}
\left[a(\Phi_N ^{\dag} \Phi_N)^2+b \Phi_N ^{\dag} \Phi_N\right]\right\}},
\label{a.15}
\end{eqnarray}
where the integration is performed over all $(N+1)\times
(N+1)$ complex matrices $\Phi_N$. It is well known
\cite{morris} that the invariance of the subintegrals in eq.(\ref{a.15})
under the transformation $\Phi_N\rightarrow U_1\Phi_N U_2$
of the double unitary group $U(N+1)\times U(N+1)$ leads to the
equality between eq.(\ref{a.14}) and eq.(\ref{a.15}).

Note also that the same steps hold if the monic classical polynomials
$P_n$ replace ${\cal P}_n$ in the above equations, then the
equality of (\ref{a.12}) with (\ref{a.15}) is a special
case of evaluation of connected correlation functions of
$U(N+1)$ invariant operators \cite{gross}.

Formula (\ref{a.15}) establishes an exact correspondence between
the random matrix model with quartic potential $V(\Phi_N ^{\dag} \Phi_N) =
b \Phi_N ^{\dag}\Phi_N + a(\Phi_N ^{\dag}\Phi_N)^2$ (in
which $\Phi_N$ is a $(N+1)\times(N+1)$ random complex matrix
and $\Phi_N ^{\dag}$ is its adjoint)
and quasi-exactly solvable
anharmonic oscillator model with sextic potential
$W_N(x^2)=[b^2-(8-2\hbar_N)a]x^2+4abx^4+4a^2x^6$ (in which $N+1$
is the number of exactly calculable eigenvalues and
corresponding eigenfunctions).

\section{Rectangular random matrices}

Let us now consider another class of quasi-exactly solvable models
described by the hamiltonian
\begin{eqnarray}
H_N=-\hbar_N^2\frac{\partial^2}{\partial x^2}+\hbar_N^2
\frac{(4s-1)(4s-3)}{4x^2}+
\{b^2-[8+(8s-4)\hbar_N]a\}x^2+4abx^4+4a^2x^6
\label{bb.1}
\end{eqnarray}
and defined on the positive half-axis $x\in[0,\infty]$.
These models differ from the models (\ref{a.1}) by the
presence of an additional parameter $s$. If $s=1/4$ then (\ref{bb.1})
reduce to (\ref{a.1}). The exactly calculable wavefunctions
in models (\ref{bb.1}) have the form
\begin{eqnarray}
\Psi_n(x)=
{\cal P}_n(x^2)(x^2)^{s-1/4}\exp\left\{-\frac{ax^4+bx^2}{2\hbar_N}\right\},
\label{b.6}
\end{eqnarray}
where ${\cal P}_n(x^2)$ are polynomials
orthogonal with the weight $|x|^{4s-1} \exp[-(ax^4+bx^2)/\hbar_N]$.
Note however that the weight function contains now an
additional factor $|x|^{4s-1}$, and therefore, instead of
formula (\ref{a.14}), we obtain
\begin{eqnarray}
\mbox{Tr}_N f(x^2)=\qquad\qquad\qquad\qquad\qquad\qquad\qquad\qquad
\nonumber\\
=\frac{\int_{0}^{+\infty}
\prod_{i=1}^{N+1}x_i^{4s-1}dx_i \left\{\sum_{i=1}^{N+1}
f(x_i^2)\right\} \prod_{i\neq k}^{N+1}(x_i^2-x_k^2)
\exp\left\{-\hbar_N^{-1}\sum_{i=1}^{N+1} (ax^4_i+bx^2_i)\right\}}
{\int_{0}^{+\infty} \prod_{i=1}^{N+1}x_i^{4s-1}dx_i
\prod_{i\neq k}^{N+1}(x_i^2-x_k^2)
\exp\left\{-\hbar_N^{-1}\sum_{i=1}^{N+1} (ax^4_i+bx^2_i)\right\}}.
\label{b.14}
\end{eqnarray}
There are two ways
of transforming this expression into the fraction of
matrix integrals. The first way is direct. We
rewrite the product $\prod_{i=1}^{N+1}x_i^{4s-1}$ as
$\exp\left\{(2s-1/2) \sum_{i=1}^{N+1}\ln x_i^2\right\}$
and obtain the random matrix theory with a non-polynomial
potential $V(\Phi_N ^{\dag}\Phi_N) = -(2s-1/2)\ln(\Phi_N ^{\dag} \Phi_N) +
\hbar_N^{-1}[b\Phi_N ^{\dag}\Phi_N + a(\Phi_N ^{\dag}\Phi_N)^2]$.
The second way is more
interesting. It is based on the observation that the
expression (\ref{b.14}) naturally arises in the model of
random complex rectangular matrices $\Phi_N^s$ of the size
$(N+1)\times4s(N+1)$ with quartic potential
$V(\bar\Phi_N^s\Phi_N^s) = \hbar_N^{-1}[b\bar\Phi_N^s\Phi_N^s +
a(\bar\Phi_N^s\Phi_N^s)^2]$ (and without any singular term!).Here
$ \bar\Phi_N^s $ is the transpose of the conjugate of the
matrix $\Phi_N^s $.
This enables one to assert that relation (\ref{a.15}) holds
for all quasi-exactly solvable models of the form (\ref{bb.1})
but the integration in the right-hand side is generally performed
over all complex rectangular matrices of size
$(N+1)\times4s(N+1)$ \cite{barb,cic,sim,periwal}.

\section{The large $N$ limit}

Up to now we considered $N$ as an arbitrary non-negative integer. Now
consider the case when $N\rightarrow\infty$. In large $N$
limit the models (\ref{a.1}) and (\ref{bb.1}) cease to be quasi-exactly
solvable and become exactly non-solvable. The
$N$-dependence of their potentials disappears and they take the
form
\begin{eqnarray}
W(x^2)=(b^2-8a)x^2+4ab x^4+4a^2 x^6.
\label{a.17}
\end{eqnarray}
The quantized Planck constant $\hbar_N$ tends to zero and
we obtain a typical semi-classical situation for all the
spectrum of the model (\ref{a.17}). In practice, however,
we do not need the information of all the states because the
left hand side of formula (\ref{a.15}) requires the knowledge
of only first $N+1$ even excitations. For them we
can use the standard semi-classical expressions.
Substituting these expressions into formula (\ref{a.16}) we
obtain
\begin{eqnarray}
\mbox{Tr}_N f(x^2)=\sum_{n=0}^{N}\frac{\oint
\frac{f(x^2)d x}{\sqrt{E_{2n}-W(x^2)}}}
{\oint
\frac{d x}{\sqrt{E_{2n}-W(x^2)}}}.
\label{a.18}
\end{eqnarray}
where $E_{2n}$ can be found from the Bohr quantization rule
\begin{eqnarray}
\frac{1}{2\pi}
\oint dx\sqrt{E_{2n}-W(x^2)} = \hbar_N \left(2n+\frac{1}{2}\right).
\label{a.19}
\end{eqnarray}
Comparing (\ref{a.15}) with (\ref{a.18}) we arrive at the
final relation
\begin{eqnarray}
\frac{\int D[\Phi_N ^{\dag},\Phi_N]
\left\{\mbox{Tr} f(\Phi_N ^{\dag}\Phi_N)\right\}
\exp\left\{-\frac{1}{\hbar_N}\mbox{Tr}
\left[a(\Phi_N ^{\dag}\Phi_N)^2+b \Phi_N ^{\dag}\Phi_N\right]\right\}}
{\int D[\Phi_N ^{\dag},\Phi_N]
\exp\left\{-\frac{1}{\hbar_N}\mbox{Tr}
\left[a(\Phi_N ^{\dag}\Phi_N)^2+b \Phi_N ^{\dag}\Phi_N\right]\right\}}
=\sum_{n=0}^{N}\frac{\oint
\frac{f(x^2)d x}{\sqrt{E_{2n}-W(x^2)}}}
{\oint
\frac{d x}{\sqrt{E_{2n}-W(x^2)}}}
\label{a.20}
\end{eqnarray}
which can be considered as an
alternative tool for doing calculations in the theory of large random
matrices.
It is worth stressing that eqs.(\ref{a.19}), (\ref{a.20}) correctly describe
only the leading term of $ 1/N $ expansion for random matrix model under
consideration. In order to obtain the non-leading terms, one should take
into account the corrections to the main semi-classical approximations. It is
interesting that, if the function $ f(x^2) $ is a polynomial, than the
computation of several terms in the $ 1/N $ expansion may be performed
analytically. Indeed, for a polynomial $ f(x^2) $, the main contribution
to the right hand side of eq.(\ref{a.20}) comes from large values of $n$,
for which the analysis of the various elliptic integrals, typically appearing
in semi-classical expansions, becomes trivial. Example of such computations
will be provided in a separate paper.

\section{The general case}

It is known that, after an appropriate change of the
initial variable $x$ by a new variable $t=t(x)$,
the hamiltonian
\begin{eqnarray}
H_N=-\frac{\partial^2}{\partial x^2}+W_N(x)
\label{a.23}
\end{eqnarray}
of any one-dimensional quasi-exactly
solvable model can be represented in the form
\begin{eqnarray}
H_N=e^{-V_0(t)/2}\{C_{\alpha\beta}S_t^\alpha S_t^\beta+
C_\alpha S_t^\alpha +C\}e^{V_0(t)/2},
\label{a.230}
\end{eqnarray}
where $S_t^\alpha, \ \alpha=-,0,+$ are
first order differential operators
\begin{eqnarray}
S_t^-=\frac{\partial}{\partial t}, \qquad
S_t^0=t\frac{\partial}{\partial t}-\frac{N}{2}, \qquad
S_t+=t^2\frac{\partial}{\partial t}-Nt,
\label{a.23a}
\end{eqnarray}
realizing a $(N+1)$-dimensional representation of algebra $sl(2)$
in the space of polynomials of order $N$. The
basis of this representation is formed by monomials $\{1,t,\ldots,t^N\}$.
Formula (\ref{a.23}) means that, up to equivalence transformation,
the hamiltonian of quasi-exactly solvable model is an
element of the universal enveloping algebra of algebra
$sl(2)$. For this reason, the linear span of functions $t^n
e^{-V_0(t)/2}, \ n=0,1,\ldots, N$ is an invariant
$(N+1)$-dimensional subspace ${\cal H}_N$ of
Hilbert space and therefore the general solution of
Schr\"odinger equation for $H_N$ in ${\cal H}_N$ can be
represented in the form $\Psi_n(x)=P_n(t(x))e^{-V_0(t(x))/2},
\ n=0,1,\ldots, N$, where $P_n(t)$ are certain polynomials of degree
$N$. Because of the orthonormalizability of wavefunctions $\Psi_n(x)$,
the polynomials $P_n(t)$ are orthogonal
with the weight $e^{-V(t)}=(dx/dt)e^{-V_0(t)}$ which
enables one to repeat the reasonings given above (see
formulas (\ref{a.16}) -- (\ref{a.15})) and derive the
generalized analog of formula (\ref{a.15})
\begin{eqnarray}
\mbox{Tr}_N f(x)=
\frac{\int D\phi \left\{\mbox{Tr} f(\phi)\right\}
\exp[-\mbox{Tr}\ V(\phi)]}
{\int D\phi \exp[-\mbox{Tr}\ V(\phi)]}
\label{a.23b}
\end{eqnarray}
in which $\phi$ is assumed to be a random hermitian
$(N+1)\times (N+1)$ matrix. Sometimes it is convenient to
represent hermitean matrix $\phi$ in the form $ \Phi ^{\dag}\Phi$ where
$\Phi$ is an arbitrary complex matrix and $\Phi ^{\dag}$ is its
adjoint. Then we arrive at direct
generalizations of formula (\ref{a.15}).

\section{Random matrix theory, virial theorems and $sl(2)$ algebra}

Assuming that $N$ is finite
consider function $f(x)$ of the form
\begin{eqnarray}
f(x)=W_N(x)+\frac{1}{2}xW'_N(x).
\label{a.21}
\end{eqnarray}
Then, according to the well known virial theorem, the left
hand side of formula (\ref{a.23b}) becomes
\begin{eqnarray}
\mbox{Tr}_N f(x)= \mbox{Tr}_N H_N,
\label{a.22}
\end{eqnarray}
where $H_N$ is the hamiltonian of the model (\ref{a.230}).
But from (\ref{a.23}) it immediately follows that
\begin{eqnarray}
\mbox{Tr}_N f(x)= C_{\alpha\beta}g^{\alpha\beta}+(N+1)C,
\label{a.25}
\end{eqnarray}
where $g^{\alpha\beta}$ is the Killing -- Cartan tensor of
algebra $sl(2)$. Therefore,
\begin{eqnarray}
\frac{\int D\phi \left\{\mbox{Tr} [W_N(\phi)+
\phi W'_N(\phi)]\right\}
\exp[-\mbox{Tr}\ V(\phi)]}
{\int D\phi \exp[-\mbox{Tr}\ V(\phi)]}=
C^{\alpha\beta}g_{\alpha\beta}+(N+1)C.
\label{a.26}
\end{eqnarray}
Thus we have found a purely Lie algebraic expression for
random matrix integrals associated with quasi-exactly
solvable models.

\section{Conclusion}

We have completed the exposition of our approach to
random matrix models associated with quasi-exactly solvable
problems in quantum mechanics. We see that these models are
actually simpler than general random matrix models and can be
solved in a systematic way by constructing the
semiclassical expansions for the associated quantum problems.
Since the parameter of the semiclassical expansion --- the quantized
Planck constant is given by the formula $\hbar_N=1/(N+1)$,
this expansion is equivalent to the topological expansion
(1/N expansion) in random matrix models. This fact, which
we consider as the main result of the present paper,
provides a new way for doing calculations in quasi-exactly
solvable random matrix models. It is worth stressing in this connection
that these models are not some exotic `monsters' but are
rather simple and ordinary looking and are rather often
discussed in the literature.

In conclusion, we would like to stress that in this paper
we did not intend to present some systematic calculations
of observables in quasi-exactly solvable random matrix models.
Our aim was only to describe a general scheme, as to the applications,
we leave them for the fortcoming publications (see e.g. ref. \cite{cicstrush}).

\section{Aknowledgements}

One of us (AGU) is grateful to Prof. Dieter Mayer for interesting
discssions. He also  thanks the staff of Theoretical
Physics Department of the University of Parma for kind
hospitality.

\end{document}